\documentclass[11pt]{article}
\usepackage{hyperref}
\usepackage{amsmath}
\usepackage{amssymb}
\usepackage{bm}
\usepackage{latexsym}
\usepackage{graphicx}
\usepackage{euscript} 
\usepackage{mathrsfs}
\begin{document}
\title{Slow-roll inflationary scenario in the maximally extended background}
\author{Ali A. Asgari \\Amir H. Abbassi*\\
\\\vspace{6pt} Department of Physics, School of Sciences,\\ Tarbiat Modares University, P.O.Box 14155-4838, Tehran, Iran\\}
\maketitle
\begin{abstract}
During the inflationary epoch, geometry of the universe may be described by quasi-de Sitter space. On the other hand, maximally extended de Sitter metric in the comoving coordinates accords with a special FLRW model with positive spatial curvature, so in this article we focus on the positively curved inflationary paradigm. For this purpose, first, we derive the power spectra of comoving curvature perturbation and primordial gravitational waves in a positively curved FLRW universe according to the slowly rolling inflationary scenario. It can be shown that the curvature spectral index in this model automatically has a small negative running parameter which is compatible with observational measurements. Then, by taking into account the curvature factor, we investigate the relative amplitude of the scalar and tensor perturbations. It would be clarified that the tensor-scalar ratio for this model against the spatially flat one, depends on the wavelength of the perturbative modes directly.  
\end{abstract}
\section{Introduction}
\label{sec:intro}
Inflationary cosmology which was proposed in the early 1980's, extends the standard Big-Bang model by postulating an early epoch of nearly exponential expansion in order to resolve a number of puzzles of the Big-Bang cosmology such as flatness, horizon and monopole problems \cite{ref1,ref2,ref3}. Inflation also explains the origin of the CMB anisotropies and large scale structure of the cosmos, indeed quantum vacuum fluctuations of the inflation field(s) magnified to the cosmic sized classical perturbations after the horizon exit time and became the seeds for the growth of the structure and CMB anisotropies in the universe\cite{ref4,ref5,ref6,ref7}. Before the advent of inflation the initial perturbations were \textit{postulated} and their spectrum was supposed to be scalar-invariant in order to fit observational data\cite{ref8,ref9,ref10}. On the other hand, inflationary theory not only truly explains the origin of the primordial inhomogeneities but also predicts their spectrum. The spectrum of these inhomogeneities, as well as the spectrum of cosmological gravitational waves produced during the inflation, are about the only observational test of the inflationary theories. Cosmological observations are consistent with the simplest model of inflation within the slow-roll paradigm \cite{ref11,ref12}. According to this scenario, the curvature power spectrum is nearly flat\cite{ref13,ref14,ref15,ref16,ref17,ref18,ref19}i.e.
\begin{equation}\label{1}
\mathcal{R}^o_q\propto q^{-\frac{3}{2}-2\epsilon-\delta},
\end{equation}
where $ \mathcal{R}_q $ is the Fourier component of comoving curvature perturbation with comoving wave number $ q $ (the superscript "o" is standing for "outside the Hubble horizon"). Furthermore, $ \epsilon $ and $ \delta $ are respectively first and second slow-roll parameters. According to the observational data $ \epsilon \leq 0.008 $ and $ \delta\leq 0.018 $ \cite{ref11}. $\mathcal{R}$ characterizes the adiabatic scalar perturbations which for super-Hubble scales has a constant value   \cite{ref20,ref21}. On the other hand, all inflationary models predict the existence of cosmological gravitational waves which produce a B-mode polarization pattern in the CMB anisotropies. Recently this mode has been detected by the BICEP2 collaboration \cite{ref22}. In the slow-roll approximation we have \cite{ref13}
\begin{equation}\label{2}
\mathcal{D}^o_q\propto q^{-\frac{3}{2}-\epsilon},
\end{equation}
where $ \mathcal{D}_q $ is the amplitude of inflationary gravitational waves. The relative amplitude which is characterized by the tensor-scalar ratio $ r=4\vert\frac{\mathcal{D}^o_q}{\mathcal{R}^o_q}\vert^2 $ is a probe of the energy scale in the inflationary epoch. It can be shown that in slow-roll approximation with a single scalar field $ r=16\epsilon $ \cite{ref13}. The BICEP2 collaboration has reported $ r\simeq 0.2 $ which is greater than the upper limit $ r<0.11 $ obtained by the Planck collaboration \cite{ref11}. In addition to this inconsistency, there are another discrepancy which refers to the running parameter of the curvature spectral index. In the slow-roll single field inflation the running parameter is of the second order in terms of slow-roll parameters, but the Planck data prefer $ \mathfrak{N}_r=\frac{\partial \mathfrak{N}_s}{\partial ln q}\simeq -0.015 $ \cite{ref11} which is of the first order slow-roll parameters and so has no any justification in slow-rolling inflationary model. In other words, the running parameter has magnitude significantly greater then the slow-roll paradigm prediction in spatially flat inflationary universe. On the other hand, there are some anomalies in the CMB power spectrum such as suppression of the lowest CMB multiples \cite{ref23} and lack of temperature correlations on scales beyond $ 70^\circ $ \cite{ref24} which may be evidences for discrete spectrum and non-trivial spherical topologies \cite{ref25,ref23}. In other words, some positive curvature models with non-trivial topology can solve the problem of the CMB quadrupole and octopole suppression and also mystery of missing fluctuations which apear in the concordance model \cite{ref26,ref27}. Furthermore, as we know the inflationary universe background is described by quasi-de Sitter space. However, maximally extended de Sitter space which is known as the \textit{Lorentzian de Sitter space} is included in the FLRW models with $ K=+1 $ and so has the positive spatial curvature \cite{ref28}. Lorentzian de Sitter is geodesically complete too. Besides, the last observational data does not roll out $ \Omega_K<0 $ case as well \cite{ref29}. It is noticeable that if the spatial curvature of the universe is positive then, the curvature dominates at early times in inflationary era\cite{ref25,ref30} , so the curvature may be significant in primordial spectra of perturbations and cannot be ignored. Dynamics of the inflationary universe with positive spatial curvature has been studied by Ellis and his collaborators\cite{ref30,ref31}. They showed that whatever the number of e-foldings increases, the curvature parameter decreases and the universe would be closer to flat today\cite{ref30}. On the other hand, Vilenkin discussed a cosmological model in which the inflationary universe is created by a quantum tunneling from \textit{nothing}\cite{ref32,ref33,ref34}. This model doesn’t have Big-Bang singularity and predicts that the inflationary universe is positively curved. Although Linde has claimed that it is very difficult to obtain a realistic model of a closed inflationary universe\cite{ref35}, Ellis and Maartens has constructed a single field inflationary model in the closed universe which is known as the \textit{eternal emergent universe scenario}\cite{ref36,ref37}. This model is a nonsingular closed inflationary cosmology that begins from a meta-stable Einstein static state. Another closed inflationary model with positive curvature index has been introduced by Lasenby and Doran\cite{ref38}.\\
In this article we investigate the slow-rolling inflationary scenario in a spatially closed background with trivial topology namely positively curved FLRW universe. The layout of the article is as follows: In section 2, we derive the generalized Mukhanov-Sasaki equation associated to the positively curved universe. Then, the slow-roll parameters and also Sasaki-Mukhanov variable is generalized to the inflationary universe with positive curvature index. Section 2 is concluded with the calculation of the comoving power spectrum. Section 3 is allotted the investigation of gravitational wave spectrum in the positively curved universe and section 4 includes the calculation of the tensor-scalar ratio in the FLRW universe with positive curvature index. Conclusion is given in the final section.  

\section{Curvature power spectrum in the positively curved FLRW universe }
\label{sec:2}
\subsection{the Mukhanov-Sasaki equation associated with the positively curved inflationary universe}
\label{sec:2.1}
In order to find the curvature power spectrum in a spatially closed universe, we should generalize the ordinary Mukhanov-Sasaki equation \cite{ref7,ref13,ref39,ref40} to the case $ K=+1 $ ($ K $  is the curvature index in the FLRW metric). This equation describes the evolution of the comoving curvature perturbation in the inflationary epoch. For this purpose, we suppose the homogeneous inflation field $ \bar{\Phi}\left( t\right)  $ has been perturbed by a small fluctuations $  \delta \Phi\left(t,\mathbf{x} \right) $ during the inflation era (hereafter bar over any quantity stands for its unperturbed value). This fluctuations are accompanied by the (scalar) perturbation in the FLRW metric (with $ K=+1 $ ). According to this perturbation, the line element of the universe may be written as\cite{ref13} 
\begin{align}
&ds^2 =  - \left( {1 + E} \right)dt^2 + 2a\left( \partial _i F \right)dt d{x^i} +a^2 \left(1 + 2\mathcal{R} \right)\tilde{ g}_{ij}dx^idx^j ,\label{3}\\ 
&\tilde{g}_{ij}=\delta_{ij}+\frac{x^ix^j}{1-\mathbf{x}^2},\nonumber
\end{align}
which is the FLRW metric with $ K=+1 $ in the comoving quasi-Cartesian coordinates $ x^{i} $ plus the scalar linear perturbation in the \textit{comoving gauge}. Here $ E $ and $ \mathcal{R} $ are respectively the \textit{lapse function} and \textit{comoving curvature perturbation}. It can be shown that in the comoving gauge\cite{ref13}
\begin{eqnarray}
&&\delta{\rho} = \delta p =- \frac{1}{2}E\dot{\bar{\Phi}}^2,\label{4}\\
&&\delta{\Phi} = 0,\label{5}
\end{eqnarray}
where $ \rho $ and $ p $ are energy density and pressure of the perfect fluid associated to the inflaton (dot stands for the derivation respect to the cosmic time $ t $ ). On the other hand, according to the perturbative field equations as well as the energy conservation law $ E $ , $ F $  and $ \mathcal{R} $ don’t evolve independently and we have\cite{ref41}
\begin{align}
& \frac{4\mathcal{R}}{a^2}+Ha\nabla ^2 F-6H \dot{\mathcal{R}}-\ddot{\mathcal{R}}+\left(3H^2+\dot{H} \right)E+\frac{1}{2}H\dot{E}+\nabla ^2 \mathcal{R}=0,\label{6} \\
&2\dot{\mathcal{R}}-HE+\frac{2}{a}F=0,\label{7} \\
& 4HaF+2a\dot{F}+E+2\mathcal{R}=0,\label{8}\\
&\dot{\delta\rho}-\dot{\bar{\Phi}}^2\left(a\nabla ^2 F-3\dot{\mathcal{R}}+3H E \right)=0,\label{9}
\end{align}
where equation (\ref{9}) is the energy conservation law. Notice that $ \nabla ^2 =\frac{1}{a^2}\tilde{g}^{ij}\nabla _i\nabla_j $ is the \textit{Laplace-Beltrami operator} respect to the $ a^2\tilde{g}_{ij} $. After some tedious and lengthy calculation, we may combine these equations and extract an explicit equation in terms of $ \mathcal{R} $ 
\begin{align}
&\Bigg[Ha^2\left(n^2-4 \right)+\frac{1}{H}-\frac{\dot{H}a^2}{H} \Bigg] \ddot{\mathcal{R}}_\mathbf{n}+\Bigg[Ha^2\frac{\dot{\boldsymbol{\chi}}}{\boldsymbol{\chi}}\left(n^2-4 \right)-\dot{H}a^2 \left(2n^2-5 \right)\nonumber\\
& \qquad +3H^2a^2\left(n^2-4 \right)+3\Bigg] \dot{\mathcal{R}}_\mathbf{n}+\Bigg[H\left(n^2-4 \right)\left(n^2-5 \right)+\frac{\dot{H}}{H}\left(n^2-3\right)\nonumber\\
&\qquad +\frac{1}{Ha^2}\left(n^2-5\right)-\frac{\dot{\boldsymbol{\chi}}}{\boldsymbol{\chi}}\left(n^2-4 \right) \Bigg] \mathcal{R}_\mathbf{n}=0, \label{10}
\end{align}
where $ \boldsymbol{\chi}=\dot{H}-\frac{1}{a^2} $ and $ \mathcal{R}_\mathbf{n} $ is the Fourier component of $ \mathcal{R} $ with comoving canonical wave number $ n $. Notice that  $ \mathbf{n}=\left(n,l,m \right) $ where $ n=3,4,..., 0\leq l\leq n-1 $ and $ \vert m\vert\leq l $ \cite{ref41}. Here due to the compactness of spatial section of spacetime, the comoving wave number is discrete. Furthermore, wave numbers $ n=1,2 $ correspond to the pure gauge \cite{ref42,ref43}, so we ignore them. One can rewrite equation (\ref{10}) in terms of conformal time $ \tau $
\begin{align}
&\Bigg[\left(n^2 - 3 \right)\mathcal{H} + \frac{1}{\mathcal{H}} - \frac{\mathcal{H}'}{\mathcal{H}} \Bigg]\mathcal{R}''_\mathbf{n} + \Bigg[2\left(n^2 - 3 \right)\mathcal{H}^2-  2\left(n^2 - 3 \right)\mathcal{H}' + \left(n^2 - 4 \right)\mathcal{H}\frac{\chi '}{\chi} + 2 \Bigg]\mathcal{R}'_\mathbf{n} \nonumber\\
&+ \Bigg[ \left(n^2-3\right)\left( n^2-5\right) \mathcal{H} + \left( n^2 - 3 \right)\frac{\mathcal{H}'}{\mathcal{H}} + \left( n^2 - 5 \right)\frac{1}{\mathcal{H}} - \left( n^2 - 4 \right)\frac{\chi '}{\chi} \Bigg]\mathcal{R}_\mathbf{n} = 0,\label{11}
\end{align} 
where the prime symbol indicates derivation respect to the conformal time. Moreover, $ \mathcal{H}=Ha $  is the comoving Hubble parameter and $ \chi=\mathcal{H}^2-\mathcal{H}'+1=4\pi G \bar{\phi}'^2 $ (indeed $  \boldsymbol{\chi}=-\frac{\chi}{a^2} $). Equation (\ref{11}) is the \textit{generalized Mukhanov-Sasaki equation} for the inflationary universe with positive curvature index.
\subsection{Re-definition of the slow-roll parameters; generalized Sasaki-Mukhanov variable}
\label{sec:2.2}
Now let’s consider the slow-roll inflation which guarantees slowly variation of inflaton by considering a Coleman–Weinberg type potential. In general, slow-roll inflation may be described by two flatness conditions\cite{ref7,ref13}
\begin{align}
&\dot{\bar{\phi}}^2\ll  V\left( \bar{\phi} \right),\label{12}\\
&\vert \ddot{\bar{\phi}}\vert \ll H\vert \dot{\bar{\phi}}\vert . \label{13} 
\end{align}
In spatially flat case relations (\ref{12}) and (\ref{13}) reduce to 
\begin{align}
&\epsilon:=-\frac{\dot{H}}{H^2}\ll 1,\label{14}\\
&\delta:=\frac{\ddot{H}}{2H\dot{H}}\ll 1\label{15}.
\end{align}
Here $ \epsilon $ and $ \delta $ are respectively first and second slow-roll parameters which are considered roughly constant. On the other hand, for the positively curved inflationary universe, the flatness conditions may be written as the same as equations (\ref{14}) and (\ref{15}) by re-definition of the slow-roll parameters
\begin{align}
&\epsilon:=-\frac{\boldsymbol{\chi}}{H^2+\frac{1}{a^2}}= \frac{\mathcal{H}^2-\mathcal{H}'+1}{\mathcal{H}^2+1}\ll 1,\label{16}\\
&\delta:= \frac{1}{2H}\frac{\dot{\boldsymbol{\chi}}}{\boldsymbol{\chi}}= \frac{1}{2\mathcal{H}}\frac{\chi'}{\chi}-1\ll 1.\label{17}
\end{align}
One can re-write equation (\ref{16}) as
\begin{equation}
\left(\frac{1}{\mathcal{H}} \right) '=-\left(1- \epsilon\right) \left(1+\frac{1}{\mathcal{H}^2} \right), 
\end{equation}
which results in
\begin{equation}\label{18}
\mathcal{H} =  - \cot \Big[\left( 1 - \epsilon \right)\tau  - \cot ^{ - 1}n\Big].
\end{equation}
Here we assumed $ \tau=\tau_n=-\int_t^{t_n}\frac{d\eta}{a\left(\eta\right)} $ where $ t_n $ is the horizon exit time for the inhomogeneity mode $ n $ ($ n=\mathcal{H}\left( t_n\right)  $). Furthermore, combination of equations (\ref{17}) and (\ref{18}) results in
\begin{equation}\label{19}
\frac{\chi '}{\chi}=-2\left( 1+\delta \right) \cot \Theta ,
\end{equation}
where $ \Theta=\left(1 - \epsilon \right)\tau  - \cot ^{ - 1}n  $.
Now by substituting of equations (\ref{18}) and (\ref{19}) in equation (\ref{11}) we can deduce 
\begin{align}
&\Bigg[\left( n^2 - 4 \right) + \frac{\epsilon }{\cos ^2\Theta} \Bigg]\mathcal{R}''_\mathbf{n} - \Bigg[ 4\left( n^2 - 4\right)\cot 2 \Theta + 4\epsilon \frac{n^2 - 3}{\sin 2 \Theta } + 2\delta\left(n^2 - 4\right) \cot \Theta \Bigg]\mathcal{R}'_\mathbf{n}+\nonumber\\ 
&\quad \Bigg[\left(n^2 - 4\right)\left(n^2 - 5\right) + 2\left(n^2 - 4 \right)\tan ^2 \Theta  - \epsilon \frac{n^2 - 3}{\cos ^2 \Theta} - 2\delta \left( n^2 - 4\right) \Bigg]\mathcal{R}_\mathbf{n} = 0 .\label{20}
\end{align}
Hereafter, we just investigate linear perturbations i.e. the terms such as $ \epsilon^2,\delta^2,\epsilon\delta , $  etc. shall be ignored. Now let’s define the new variable $ \mathcal{V}_\mathbf{n} $ as
\begin{equation}\label{21}
\mathcal{V}_\mathbf{n} =\mathscr{T}\mathcal{R}_\mathbf{n}, \quad \mathscr{T}=\mathscr{C}\frac{\exp\left[ - \frac{\epsilon }{2\left(n^2 - 4 \right)\cos ^2 \Theta }\right]} {\left| \sin\Theta\right|^{1 + 2\epsilon  + \delta }\left|\cos  \Theta \right|}.
\end{equation}
(Here $ \mathscr{C} $ is a constant which is obtained soon) Thus, equation (\ref{20}) may be written in terms of $ \mathcal{V}_\mathbf{n} $
\begin{multline}\label{22}
\mathcal{V}''_\mathbf{n} + \Bigg[ \left( n^2 - 5 \right) - 2\cot ^2\Theta + \epsilon \left(1 + \cot ^2 \Theta\right)\left( 2\frac{1 - \cot ^2 \Theta}{\cot ^2 \Theta} + \frac{1}{n^2 - 4}\frac{3 - \cot ^2 \Theta}{\cot ^4\Theta} \right)\\ - \delta \left(1 + 3\cot ^2 \Theta \right) \Bigg]\mathcal{V}_\mathbf{n} = 0 .
\end{multline}
Before solving equation (\ref{22}), let’s find constant $  \mathscr{C} $. For this purpose, we may invoke the relation
\begin{equation}\label{23}
\frac{\epsilon '}{\epsilon}=2\mathcal{H}\left(\epsilon +\delta \right), 
\end{equation}
which can be derived from the logarithmic derivation of equation (\ref{16}). Provided that $ \epsilon $ and $ \delta $ are supposed to be constant, equation (\ref{23}) yields
\begin{equation}\label{24}
\left(\frac{a}{\mathfrak{R}} \right) ^{\epsilon +\delta}=\sqrt{\epsilon} .
\end{equation} 
Here $ \mathfrak{R} $ is a characteristic scale which is appeared as the integral constant in equation (\ref{24}).\\
On the other hand, equation (\ref{18}) results in
\begin{equation}\label{25}
a=\frac{1}{\mathfrak{H}}\left|\sin\Theta\right| ^{-\left(\epsilon +\delta\right) }, \quad \mathfrak{H}=\sqrt{\frac{8\pi G}{3}\bar{\rho}} ,
\end{equation}
Consequently
\begin{equation}\label{26}
\left|\sin\Theta\right| ^{\epsilon +\delta}=\frac{\left(\mathfrak{RH}\right) ^{-\left(\epsilon +\delta\right) }}{\sqrt{\epsilon}} .
\end{equation}
Meanwhile, using equations (\ref{18}), (\ref{21}) as well as (\ref{16}) and (\ref{17}) one can show
\begin{equation}\label{27}
\frac{\mathscr{T}'}{\mathscr{T}}=\frac{a'}{a}-\frac{\mathcal{H}'}{\mathcal{H}}+\frac{\bar{\phi}''}{\bar{\phi}'}+\frac{1}{n^2-4}\left(\frac{1}{\mathcal{H}}-\frac{\mathcal{H}'}{\mathcal{H}^3}+\frac{1}{\mathcal{H}^3} \right) ,
\end{equation}
which results in
\begin{equation}\label{28}
\mathscr{T}=\frac{a\bar{\phi}'}{\mathcal{H}}\exp\left[ \frac{1}{n^2-4}\left(\frac{1}{2\mathcal{H}^2}+\int\frac{\mathcal{H}^2+1} {\mathcal{H}^3} d\tau\right)\right] .
\end{equation}
Now let’s suppose $ n\longrightarrow+\infty $, thus equation (\ref{28}) takes the form
\begin{equation}\label{29}
\mathop {\lim }\limits_{n \to +\infty }\mathscr{T}=\frac{\mathscr{C}} {\left| \sin\Theta\right|^{1 + 2\epsilon  + \delta }\left|\cos  \Theta \right|}=\frac{a\bar{\phi}'}{\mathcal{H}}=\mathscr{Z} .
\end{equation}
For the severe sub-Hubble modes, the curvature has a negligible imprint and may be disregarded, so it coincides with $ K=0 $ case. Besides, it can be shown that
\begin{equation}\label{30}
\frac{a\bar{\phi}'}{\mathcal{H}}=\frac{a}{\mathcal{H}}\sqrt{\frac{\mathcal{H}^2-\mathcal{H}'+1}{4\pi G}}=a\sqrt{\left( 1+\frac{1}{\mathcal{H}^2}\right)\frac{\epsilon}{4\pi G} }=\frac{1}{\mathfrak{H}}\left|\sin\Theta\right|^{-1-\epsilon}\left|\cos\Theta\right|^{-1}\sqrt{\frac{\epsilon}{4\pi G}},
\end{equation}
Therefore, combination of equations (\ref{30}), (\ref{29}) and (\ref{26}) yields
\begin{equation}
\mathscr{C}=\frac{1}{\sqrt{4\pi G}}\frac{1}{\mathfrak{H}\left( \mathfrak{RH}\right) ^{\epsilon +\delta}} ,
\end{equation} 
So
\begin{equation}\label{31}
\mathcal{R}_\mathbf{n} =\sqrt{4\pi G}\mathfrak{H}\left( \mathfrak{RH}\right) ^{\epsilon +\delta}\left| \sin\Theta\right|^{1 + 2\epsilon  + \delta }\left|\cos  \Theta \right| \exp\left[\frac{\epsilon }{2\left(n^2 - 4 \right)\cos ^2 \Theta }\right]\mathcal{V}_\mathbf{n} .
\end{equation}
Notice that $ \mathcal{V}_\mathbf{n} $ is the \textit{generalized Sasaki-Mukhanov variable}  for the inflationary universe with positive curvature index. 
\subsection{Curvature power spectrum }
\label{sec:2.3}
Now let’s find the solutions of the equation (\ref{22}). For this purpose, we assume $ x=\cos\Theta $, thus equation (\ref{22}) reduces to
\begin{multline}\label{32}
\left(1-x^2 \right)\frac{d^2\mathcal{V}_\mathbf{n}}{dx^2}-x\frac{d\mathcal{V}_\mathbf{n}}{dx}+\\
\left[ \left(n^2-3 \right)\left(1+2\epsilon \right)+2\delta-\frac{2+6\epsilon +3\delta}{1-x^2}+2\epsilon\left(1-\frac{2}{n^2-4}\right) \frac{1}{x^2}+\frac{3\epsilon}{n^2-4}\frac{1}{x^4} \right] \mathcal{V}_\mathbf{n}=0 .
\end{multline}   
It may be proposed the solution as
\begin{equation}\label{33}
\mathcal{V}_\mathbf{n}=\EuScript{V}_\mathbf{n}+\epsilon \mathfrak{V}_\mathbf{n} ,
\end{equation}
where
\begin{equation}\label{34}
\EuScript{V}_\mathbf{n}=\mathscr{A}\sqrt[4]{1-x^2} P^\mu_\nu\left(x \right) +\mathscr{B}\sqrt[4]{1-x^2} Q^\mu_\nu\left(x \right),
\end{equation}

\begin{equation}\label{35}
\left\{
\begin{aligned}
&\mu:=\frac{3}{2}+2\epsilon+\delta ,\\
&\nu:=\left(1+\epsilon \right)\sqrt{n^2-3}+\frac{\delta}{\sqrt{n^2-3}} -\frac{1}{2}.
\end{aligned}
\right.
\end{equation}
Notice that $  P^\mu_\nu $ and $  Q^\mu_\nu $ are associated Legendre functions. By inserting ansatz (\ref{33}) in equation (\ref{32}) and neglecting higher order infinitesimal terms, one obtains a second order nonhomogeneous equation in terms of $ \mathfrak{V}_\mathbf{n} $
\begin{multline}\label{36}
\left(1-x^2 \right)\frac{d^2\mathfrak{V}_\mathbf{n}}{dx^2}-x\frac{d\mathfrak{V}_\mathbf{n}}{dx}+\left[ \left(n^2-3 \right)\left(1+2\epsilon \right)+2\delta-\frac{2+6\epsilon+3\delta}{1-x^2}\right] \mathfrak{V}_\mathbf{n}=\\
-\frac{1}{n^2-4}\left(2\frac{n^2-6}{x^2}+\frac{3}{x^4} \right) \left[ \mathscr{A}\left(1-x^2 \right)^{\frac{1}{4}} P^\mu_\nu\left(x \right) +\mathscr{B}\left(1-x^2 \right)^{\frac{1}{4}} Q^\mu_\nu\left(x \right)\right] .
\end{multline}
which has the special solution as
\begin{multline}\label{37}
\mathfrak{V}_\mathbf{n}=\frac{1}{n^2-4}\frac{\Gamma\left(\nu-\mu+1 \right) }{\Gamma\left(\nu+\mu+1 \right)}\sqrt[4]{1-x^2} \Bigg\{ \Big[ \mathscr{A}P^\mu_\nu\left(x \right)  -\mathscr{B}Q^\mu_\nu\left(x \right) \Big] \int_{x_0}^x \left(1-y^2 \right)\left(2\frac{n^2-6}{y^2}+\frac{3}{y^4} \right)\times \\
 P^\mu_\nu\left(y \right)Q^\mu_\nu\left(y \right)dy - \mathscr{A}Q^\mu_\nu\left(x \right)\int_{x_0}^x \left(1-y^2 \right)\left(2\frac{n^2-6}{y^2}+\frac{3}{y^4} \right)\Big[ P^\mu_\nu\left(y \right)\Big]^2dy  \\ +\mathscr{B}P^\mu_\nu\left(x \right)\int_{x_0}^x \left(1-y^2 \right)\left(2\frac{n^2-6}{y^2}+\frac{3}{y^4} \right)\Big[ Q^\mu_\nu\left(y \right)\Big] ^2 dy \Bigg\}  .
\end{multline}
Here, $ x_0 $ is an arbitrary constant for which $ \left| x_0\right|\leqslant 1 $. Consequently, the general solution of equation (\ref{22}) reduces to
\begin{align}\label{38}
&\mathcal{V}_\mathbf{n}\left(\tau \right) =\sqrt{\left|\sin\Theta\right|}\Big[\mathscr{A} P^\mu_\nu\left(\cos\Theta \right) +\mathscr{B} Q^\mu_\nu\left(\cos\Theta \right)\Big] + \frac{\epsilon}{n^2-4}\frac{\Gamma\left(\nu-\mu+1 \right) }{\Gamma\left(\nu+\mu+1 \right)}\sqrt{\left|\sin\Theta\right|}\times\nonumber\\
&\Bigg\{\Big[-\mathscr{A}P^\mu_\nu\left(\cos\Theta  \right) +\mathscr{B}Q^\mu_\nu\left(\cos\Theta  \right) \Big]\int_{\Theta_0}^\Theta\sin ^3\Upsilon\left(2\frac{n^2-6}{\cos ^2 \Upsilon}+\frac{3}{\cos^4\Upsilon} \right)P^\mu_\nu\left(\cos\Upsilon \right)\times\nonumber\\
&Q^\mu_\nu\left(\cos\Upsilon \right)d\Upsilon+\mathscr{A}Q^\mu_\nu\left(\cos\Theta  \right)\int_{\Theta_0}^\Theta \sin^3 \Upsilon\left(2\frac{n^2-6}{\cos^2\Upsilon}+\frac{3}{\cos^4\Upsilon} \right)\Big[ P^\mu_\nu\left(\cos\Upsilon \right)\Big]^2d\Upsilon \nonumber\\
&- \mathscr{B}P^\mu_\nu\left(\cos\Theta  \right)   \int_{\Theta_0}^\Theta\sin^3 \Upsilon\left(2\frac{n^2-6}{\cos^2\Upsilon}+\frac{3}{\cos^4\Upsilon} \right)\Big[ Q^\mu_\nu\left(\cos\Upsilon \right)\Big] ^2 d\Upsilon \Bigg\} .
\end{align}
Hereafter, we put $ \Theta_0=-\cot^{-1}n $ ( it is completely compatible with the conformal initial condition which is introduced below).\\
In order to determine the constants $  \mathscr{A} $ and $  \mathscr{B} $ we may use the \textit{conformal (Bunch-Davies) initial condition}  which states\cite{ref44,ref45}
\begin{equation}\label{39}
\mathop {\lim }\limits_{n \to +\infty }\mathcal{V}_\mathbf{n}=\frac{1}{\sqrt{2n}}\exp\left( -in\tau\right) .
\end{equation}
Thus, according to the asymptotic formulas of $  P^\mu_\nu $ and $ Q^\mu_\nu $ for large value of $  \nu $ i.e. \cite{ref46}
\begin{align}
&P^\mu_\nu\left(\cos\theta \right) \sim\frac{\Gamma\left(\mu+\nu+1 \right) }{\Gamma\left(\nu+\frac{3}{2} \right) }\sqrt{\frac{2}{\pi\sin\theta}}\sin\left[ \left(\nu+\frac{1}{2} \right)\theta+\frac{\pi}{4} +\frac{\mu\pi}{2} \right]+\mathcal{O}\left(\nu^{-1} \right) ,\label{40} \\ 
&Q^\mu_\nu\left(\cos\theta \right) \sim\frac{\Gamma\left(\mu+\nu+1 \right) }{\Gamma\left(\nu+\frac{3}{2} \right) }\sqrt{\frac{\pi}{2\sin\theta}}\cos\left[ \left(\nu+\frac{1}{2} \right)\theta+\frac{\pi}{4} +\frac{\mu\pi}{2} \right]+\mathcal{O}\left(\nu^{-1} \right) ,\label{41} 
\end{align}
and noting that $ \mathop {\lim}\limits_{n \to +\infty }\frac{\left(n+\alpha \right)!}{n!}\sim n^\alpha $, after alot of lengthy but straightforward calculations, it can be shown
\begin{equation}\label{42}
\left\{
\begin{aligned}
&\mathscr{A}=\frac{i\sqrt{\pi}}{2}n^{-\frac{3}{2}-2\epsilon-\delta},\\
&\mathscr{B}=-\frac{1}{\sqrt{\pi}}n^{-\frac{3}{2}-2\epsilon-\delta}.
\end{aligned}
\right.
\end{equation}
Thus
\begin{align}\label{43}
&\mathcal{V}_\mathbf{n}\left(\tau \right) =\sqrt{\left|\sin\Theta\right|}n^{-\mu}\Bigg\{\frac{i\sqrt{\pi}}{2} P^\mu_\nu\left(\cos\Theta \right) -\frac{1}{\sqrt{\pi}} Q^\mu_\nu\left(\cos\Theta \right)- \frac{\epsilon}{n^2-4}\frac{\Gamma\left(\nu-\mu+1 \right) }{\Gamma\left(\nu+\mu+1 \right)}\times\nonumber\\
& \Big[\frac{i\sqrt{\pi}}{2}P^\mu_\nu\left(\cos\Theta  \right) +\frac{1}{\sqrt{\pi}}Q^\mu_\nu\left(\cos\Theta  \right) \Big]\int_0^{\left( 1-\epsilon \right) \tau}\sin ^3\Upsilon\left(2\frac{n^2-6}{\cos ^2 \Upsilon}+\frac{3}{\cos^4\Upsilon} \right)P^\mu_\nu\left(\cos\Upsilon \right)\times\nonumber\\
& Q^\mu_\nu\left(\cos\Upsilon \right)d\eta+\frac{i\sqrt{\pi}}{2}\frac{\epsilon}{n^2-4}\frac{\Gamma\left(\nu-\mu+1 \right) }{\Gamma\left(\nu+\mu+1 \right)}Q^\mu_\nu\left(\cos\Theta  \right)\int_0^{\left( 1-\epsilon \right) \tau} \sin^3 \Upsilon\left(2\frac{n^2-6}{\cos^2\Upsilon}+\frac{3}{\cos^4\Upsilon} \right)\times\nonumber\\
&\Big[ P^\mu_\nu\left(\cos\Upsilon \right)\Big]^2d\eta +\frac{1}{\sqrt{\pi}}\frac{\epsilon}{n^2-4}\frac{\Gamma\left(\nu-\mu+1 \right) }{\Gamma\left(\nu+\mu+1 \right)}P^\mu_\nu\left(\cos\Theta  \right)   \int_0^{\left( 1-\epsilon \right) \tau}\sin^3 \Upsilon\left(2\frac{n^2-6}{\cos^2\Upsilon}+\frac{3}{\cos^4\Upsilon} \right)\times \nonumber\\
&\Big[ Q^\mu_\nu\left(\cos\Upsilon \right)\Big] ^2 d\eta \Bigg\} .
\end{align}
Furthermore, by ignoring the non-linear terms, $ \mathcal{R}_\mathbf{n} $ takes the form
\begin{align}\label{44}
&\mathcal{R}_\mathbf{n}\left(\tau \right) =\sqrt{4\pi G}\mathfrak{H}\left( \mathfrak{RH}\right) ^{\epsilon +\delta}\left|\frac{\sin \Xi}{n} \right|^\mu\left|\cos \Xi \right| \Big[ \frac{i\sqrt{\pi}}{2} P^\mu_\nu\left(\cos \Xi\right)-\frac{1}{\sqrt{\pi}} Q^\mu_\nu\left(\cos \Xi \right)\Big]\nonumber\\
&+\epsilon \sqrt{4\pi G}\mathfrak{H}\left( \mathfrak{RH}\right) ^{\epsilon +\delta}\left|\frac{\sin \Xi}{n} \right|^\mu\left|\cos \Xi \right| \Bigg\{ -\frac{i\sqrt{\pi}}{2}\tau\frac{dP^\mu_\nu\left(\cos\Xi \right)}{d\tau}+\frac{1}{\sqrt{\pi}}\tau\frac{dQ^\mu_\nu\left(\cos\Xi \right)}{d\tau}\nonumber\\
&-\Big[2\tau\cot 2\Xi +\frac{1}{2}\tau\cot \Xi-\frac{1}{2\left(n^2-4 \right)\cos ^2\Xi } \Big] \Big[ \frac{i\sqrt{\pi}}{2}P^\mu_\nu\left( \cos\Xi\right) -\frac{1}{\sqrt{\pi}}Q^\mu _\nu\left( \cos\Xi\right) \Big]\nonumber\\
&-\frac{1}{n^2-4}\frac{\Gamma\left(\nu-\mu+1 \right) }{\Gamma\left(\nu+\mu+1 \right)}\Big[\frac{i\sqrt{\pi}}{2}P^\mu_\nu\left(\cos\Xi  \right) +\frac{1}{\sqrt{\pi}}Q^\mu_\nu\left(\cos\Xi  \right) \Big]\nonumber\\
&\times \int_0^\tau\sin ^3\Upsilon\left(2\frac{n^2-6}{\cos ^2 \Upsilon}+\frac{3}{\cos^4\Upsilon} \right)P^\mu_\nu\left(\cos\Upsilon \right)Q^\mu_\nu\left(\cos\Upsilon \right)d\eta\nonumber\\
&+\frac{i\sqrt{\pi}}{2}\frac{1}{n^2-4}\frac{\Gamma\left(\nu-\mu+1 \right) }{\Gamma\left(\nu+\mu+1 \right)}Q^\mu_\nu\left(\cos\Xi  \right)\int_0^\tau \sin^3 \Upsilon\left(2\frac{n^2-6}{\cos^2\Upsilon}+\frac{3}{\cos^4\Upsilon} \right)\Big[ P^\mu_\nu\left(\cos\Upsilon \right)\Big]^2d\eta \nonumber\\
&+\frac{1}{\sqrt{\pi}}\frac{1}{n^2-4}\frac{\Gamma\left(\nu-\mu+1 \right) }{\Gamma\left(\nu+\mu+1 \right)}P^\mu_\nu\left(\cos\Xi  \right)   \int_0^\tau\sin^3 \Upsilon\left(2\frac{n^2-6}{\cos^2\Upsilon}+\frac{3}{\cos^4\Upsilon} \right)\Big[ Q^\mu_\nu\left(\cos\Upsilon \right)\Big] ^2 d\eta \Bigg\} ,
\end{align}
where $ \Xi= \tau-\cot^{-1}n $ and $ \Upsilon=\eta-\cot^{-1}n  $.\\
It is important to evaluate the comoving curvature perturbation at the horizon exit time $ \tau =0 $ i.e. when the quantum fluctuations of inflaton came to be classical perturbations. Besides, by inserting $  \tau =0 $ in equation (\ref{44}) the arguments of $ P^\mu_\nu $ and $ Q^\mu_\nu $ become $  \cos\left(\cot^{-1}n \right)=\frac{n}{\sqrt{n^2+1}}  $ which for $  n\geq 3 $ , $  0.94 \leq \frac{n}{\sqrt{n^2+1}}<1  $, so it may be plausible to use asymptotic formulas of the associated Legendre functions near one i.e. \cite{ref47}
\begin{align}
&\theta\longrightarrow 0\quad :\quad P^\mu_\nu\left(\cos\theta \right) \sim\frac{1}{\pi}\Gamma\left(\mu\right) \sin\mu\pi \left( \frac{2}{1-\cos\theta}\right)^{\frac{\mu}{2}} ,\label{45} \\ 
&\theta\longrightarrow 0\quad :\quad Q^\mu_\nu\left(\cos\theta \right) \sim\frac{1}{2}\Gamma\left(\mu\right) \cos\mu\pi \left( \frac{2}{1-\cos\theta}\right)^{\frac{\mu}{2}} .\label{46}
\end{align}
So by doing some straightforward calculation, it can be shown
\begin{multline}\label{47}
\mathcal{R}^o_\mathbf{n}=-\sqrt{G} \mathfrak{H}\left( \mathfrak{RH}\right) ^{\epsilon +\delta} \Gamma\left(\mu\right) \exp\left(-i\mu \pi \right)\frac{n^{1-\mu}}{\sqrt{n^2+1}}\left(2+\frac{2n}{\sqrt{n^2+1}} \right) ^{\frac{\mu}{2}}\\
\times\left(1+\frac{n^2+1}{2n^2\left(n^2-4 \right) }\epsilon\right) . 
\end{multline}
Let’s approximate $ \frac{n}{\sqrt{n^2+1}}\sim 1  $, thus equation (\ref{47}) takes the form
\begin{equation}\label{48}
\mathcal{R}^o_\mathbf{n}\simeq-\sqrt{G} \mathfrak{H}\left(\mathfrak{RH}\right) ^{\epsilon +\delta}2^{\frac{3}{2}+2\epsilon +\delta} \Gamma\left(\mu\right) \exp\left(-i\mu\pi \right)n^{-\mu}\left(1+\frac{\epsilon}{2\left(n^2-4 \right) }\right) . 
\end{equation}
Consequently, the curvature power spectrum in the maximally extended inflationary universe with single field reduces to
\begin{equation}\label{49}
\mathcal{P}^o_{\mathcal{R}}\left(n\right) \propto n^{-3-4\epsilon-2\delta}\left( 1+\frac{\epsilon}{n^2-4}\right) .
\end{equation}
Except the additional factor $  1+\frac{\epsilon}{n^2-4} $, spectrum (\ref{49}) is similar to the nearly flat spectrum which can be deduced from the slow-rolling inflationary scenario with spatially flat background\cite{ref13}.  By definition of the \textit{curvature spectral index} as 
\begin{equation}
\mathcal{P}^o_{\mathcal{R}}\left(n\right) \propto n^{\mathfrak{N}_s\left(n\right)-4} ,
\end{equation}
one can show that
\begin{equation}\label{50}
\mathfrak{N}_s\left(n\right)=1 - 4\epsilon  - 2\delta  + \frac{2\epsilon}{\left( {{n^2} - 4} \right)\ln n} .
\end{equation}
Because $ n\geq 3 $ so
\begin{equation}
1 - 4\epsilon  - 2\delta <\mathfrak{N}_s\left(n\right)\lesssim 1-3.64\epsilon-2\delta .
\end{equation}
It means the curvature spectral index in the maximally extended universe shall be a bit larger than the $ K=0 $ corresponding model (For the $ K=0 $ case, $ \mathfrak{N}_s\left(n\right)=1 - 4\epsilon  - 2\delta $). Moreover, $ \mathfrak{N}_s $ directly depends on the comoving wave number ($ n $)  and so the spectrum is running. In other words, \textit{running parameter} of $ \mathfrak{N}_s $ doesn’t vanish
\begin{equation}
\mathfrak{N}_r\left( n \right) = n\frac{\partial  \mathfrak{N}_s}{\partial n} =- 2\epsilon \frac{\left(n^2 - 4\right) + 2n^2\ln n}{\left( n^2 - 4\right)^2\ln ^2n} < 0 .
\end{equation}
It is remarkable that the sign of $ \mathfrak{N}_r $ coincides with experimental data. Moreover, running parameter in the maximally extended background inflationary model is proportional to $ \epsilon $ i.e. $ \mathfrak{N}_r $ is of the first order slow-roll parameters accordant with the reports \cite{ref11}, in spite of the spatially flat case in which against the Planck reports is roughly zero.
\section{Primordial gravitational waves power spectrum in the positively curved universe}
\label{sec:3}
The primordial gravitational waves during inflationary epoch can be treated in the same way as the comoving curvature perturbation considered in previous section. In fact, quantum fluctuations of the inflaton may result in tensorial perturbations described by a symmetric traceless divergenceless tensor field $ D_{ij}\left(t,\bf{x} \right)  $ which perturbs the FLRW metric as\cite{ref13}
\begin{equation}
ds^2=-dt^2+a^2\left(\tilde{g}_{ij}+D_{ij}\right)dx^idx^j .
\end{equation}
Propagation of $ D_{ij} $ in the positively curved FLRW universe is described by\cite{ref41}
\begin{equation}\label{51}
 a^2\nabla ^2D_{ij}-3a\dot{a}\dot{D}_{ij}-a^2\ddot{D}_{ij}-2D_{ij}=-16\pi Ga^2\Pi ^T_{ij} .
\end{equation}
Here $ \Pi ^T_{ij}\left(t,\bf{x} \right)  $ is the anisotropic inertia tensor which vanishes for the scalar fields, so
\begin{equation}\label{52}
 a^2\nabla ^2D_{ij}-3a\dot{a}\dot{D}_{ij}-a^2\ddot{D}_{ij}-2D_{ij}=0 .
\end{equation} 
One may expand $ D_{ij} $ in terms of the t-t tensor spherical harmonics on $ \mathbb{S}^3\left(a \right)  $ \cite{ref41}
\begin{equation}\label{53}
D_{ij}\left(t ,\mathbf{x} \right) = \sum\limits_{nlm} \left[ \mathcal{D}_{nlm}^{\mathbb{O}}\left(t \right)\big( T_{ij}^{\mathbb{O}} \big)_{nlm}+ \mathcal{D}_{nlm}^{\mathbb{E}}\left( t \right)\big( T_{ij}^{\mathbb{E}} \big)_{nlm} \right] ,
\end{equation}
where $ \mathcal{D}_\mathbf{n}^\mathbb{O} $ and $ \mathcal{D}_\mathbf{n}^\mathbb{E} $ correspond to two different polarizations of the gravitational waves. Notice that $ \left\lbrace \big( T_{ij}^{\mathbb{O}} \big)_{nlm} ,\big( T_{ij}^{\mathbb{E}} \big)_{nlm} \right\rbrace $ constitutes a complete orthonormal basis for the expansion of any symmetric traceless divergence-free covariant tensor field of rank 2 on $  \mathbb{S}^3\left(a \right) $. Furthermore \cite{ref41},
\begin{align}
& \nabla ^2 \big( T_{ij}^\mathbb{O}\big)_{nlm}  = \frac{3 - n^2}{a^2 }\big(T_{ij}^\mathbb{O}\big)_{nlm}, \quad n = 3,4,... \\ &
 \nabla ^2 \big(T_{ij}^\mathbb{E} \big)_{nlm}  = \frac{3 - n^2 }{a^2}\big(T_{ij}^\mathbb{E} \big)_{nlm}.\quad n = 3,4,... 
 \end{align}
Thus equation (\ref{53}) reduces to two independent equations
\begin{equation}
\left\{\begin{aligned}
&\ddot{\mathcal{D}}_\mathbf{n}^\mathbb{O}\left(t \right)  + 3H\dot{\mathcal{D}}_\mathbf{n}^\mathbb{O}\left(t \right) +\frac{n^2 - 1}{a^2} \mathcal{D}_\mathbf{n}^\mathbb{O}\left(t \right) = 0 ,\\ 
&\ddot{\mathcal{D}}_\mathbf{n}^\mathbb{E}\left(t \right) + 3H\dot{\mathcal{D}}_\mathbf{n}^\mathbb{E}\left(t \right) +\frac{n^2 - 1}{a^2} \mathcal{D}_\mathbf{n}^\mathbb{E}\left(t \right) = 0 .
\end{aligned}\right.
\end{equation}
Hereafter we omit the superscripts $ \mathbb{O} $ and $ \mathbb{E} $ because both of $ \mathcal{D}_\mathbf{n}^\mathbb{O} $ and $ \mathcal{D}_\mathbf{n}^\mathbb{E} $ satisfy the same equation 
\begin{equation}\label{54}
\ddot{\mathcal{D}}_\mathbf{n}\left(t \right)  + 3H\dot{\mathcal{D}}_\mathbf{n}\left(t \right) +\frac{n^2 - 1}{a^2} \mathcal{D}_\mathbf{n}\left(t \right) = 0 .
\end{equation}
 $ \mathcal{D}_\mathbf{n}\left(t \right)  $ is amplitude of the gravitational wave $ D_{ij}\left(t, \bf{x} \right)  $ as well as a tensor random field on $  \mathbb{S}^3\left(a \right) $. By converting the cosmic time to the conformal time equation (\ref{54}) takes the form
\begin{equation}\label{55}
\mathcal{D}''_\mathbf{n}\left(\tau \right)  + 2\mathcal{H}\mathcal{D}'_\mathbf{n}\left(\tau \right) +\left( n^2 - 1\right) \mathcal{D}_\mathbf{n}\left(\tau \right) = 0 .
\end{equation}
During the slow-rolling inflationary epoch we can write
\begin{equation}\label{56}
\mathcal{D}''_\mathbf{n}\left(\tau \right)  - 2\cot\Theta\mathcal{D}'_\mathbf{n}\left(\tau \right) +\left( n^2 - 1\right) \mathcal{D}_\mathbf{n}\left(\tau \right) = 0 .
\end{equation}
By assumption $ x=\cos\Theta $ equation (\ref{56}) can be written as
\begin{equation}
\left(1-x^2 \right)\frac{d^2\mathcal{D}_\mathbf{n}}{dx^2}+\left(1+2\epsilon \right)x\frac{d\mathcal{D}_\mathbf{n}}{dx}+\left( 1+2\epsilon\right) \left( n^2-1\right) \mathcal{D}_\mathbf{n}\left(x \right)=0 ,
\end{equation}
which has the general solution as
\begin{equation}
\mathcal{D}_\mathbf{n}\left(x \right)=\left(1-x^2 \right) ^{\frac{2\epsilon +3}{4}}\left[\mathscr{P}P^{\epsilon +\frac{3}{2}}_{n\left( 1+\epsilon\right) -\frac{1}{2}}\left(x \right)+ \mathscr{Q}Q^{\epsilon +\frac{3}{2}}_{n\left( 1+\epsilon\right) -\frac{1}{2}}\left(x \right) \right] .
\end{equation}
Here $ \mathscr{P}  $ and $ \mathscr{Q} $ are two arbitrary constants. So the solution of equation (\ref{56}) is
\begin{equation}\label{57}
\mathcal{D}_\mathbf{n}\left(\tau \right)=\left|\sin\Theta \right| ^\iota\Big[\mathscr{P}P^\iota_\kappa\left(\cos\Theta \right)+ \mathscr{Q}Q^\iota_\kappa\left(\cos\Theta \right) \Big] ,
\end{equation}
where
\begin{equation}
\left\{
\begin{aligned}
&\iota :=\epsilon +\frac{3}{2} ,\\
&\kappa :=n\left( 1+\epsilon\right) -\frac{1}{2} .
\end{aligned}
\right.
\end{equation}
Besides, the initial condition must be satisfied by $ \mathcal{D}_\mathbf{n} $  is very similar to the Bunch-Davies initial condition exerted to the Sasaki-Mukhanov variable\cite{ref13}
\begin{equation}\label{58}
\mathop {\lim }\limits_{n \to +\infty }\mathcal{D}_\mathbf{n}=\frac{\sqrt{16\pi G}}{a\left(t \right) }\frac{1}{\sqrt{2n}}\exp\left( -in\tau\right) ,
\end{equation} 
which is applicable for both polarization modes distinctly. By considering the asymptotic formulas (\ref{40}) and (\ref{41}) and equation (\ref{25}) as well, one can obtain
\begin{equation}
\left\{
\begin{aligned}
&\mathscr{P}=2\pi i \mathfrak{H}\sqrt{G}n^{-\iota} ,\\
&\mathscr{Q}=-4\mathfrak{H}\sqrt{G}n^{-\iota} .
\end{aligned}
\right.
\end{equation}
Thus
\begin{equation}\label{59}
\mathcal{D}_\mathbf{n}\left( \tau  \right) = 2\sqrt{G} \mathfrak{H}\left| \frac{\sin \Theta }{n}\right|^\iota \Big[\pi i P_\kappa ^\iota \left(\cos \Theta\right) - 2Q_\kappa ^\iota \left(\cos \Theta\right)\Big] .
\end{equation}
$ \mathcal{D}_\mathbf{n}^o $ may be determined by considering $ \mathcal{D}_\mathbf{n} $ at the time of horizon crossing ($ \tau=0 $)
\begin{equation}\label{60}
\mathcal{D}_\mathbf{n}^o=-2\sqrt{G}\mathfrak{H}\Gamma\left( \iota\right) \exp\left(-i \iota \pi \right) n^{-\iota}\left(2+\frac{2n}{\sqrt{n^2+1}} \right) ^{\frac{\iota}{2}} .
\end{equation}
Here again we use the asymptotic relations (\ref{45}) and (\ref{46}). By approximation $  \frac{n}{\sqrt{n^2+1}}\sim 1  $, equation (\ref{60}) acquires a simpler form which results in
\begin{equation}
\mathcal{P}^o_\mathcal{D}\left(n \right) \propto n^{-3-2\epsilon},
\end{equation}
So by definition of the \textit{tensor spectral index} as
\begin{equation}
\mathcal{P}^o_\mathcal{D}\propto n^{\mathfrak{N}_T-3},
\end{equation}
one can obtain
\begin{equation}
\mathfrak{N}_T=-2\epsilon ,
\end{equation}
which is perfectly analogous to tensor spectral index derived in the classical slow-rolling inflationary theory \cite{ref13}.
\section{Tensor-scalar ratio in the positively curved universe}
\label{sec:4}
\textit{Tensor-scalar ratio} in the positively curved FLRW universe may be defined as \cite{ref13}
\begin{equation}
r_n:=4\frac{\mathcal{P}_\mathcal{D}^o\left( n \right)}{\mathcal{P}_\mathcal{R}^o\left( n \right)} = 4\left|\frac{\mathcal{D}_\mathbf{n}^o}{\mathcal{R}_\mathbf{n}^o} \right| ^2 .
\end{equation}
Here the factor 4 refers to two different polarization modes of the gravitations waves. Significance of $ r_n $ come from its measurability, indeed tensor-scalar ratio can provide an assay for the inflationary scenarios and some inflation theories may be crossed out due to the contradiction with observational value of $ r_n $.  According to the standard slow-rolling inflationary theory $ r_q=16\epsilon $ ($ q $ stands for the comoving wave number of perturbations in the spatially flat universe) \cite{ref13}, so if one suppose $ \epsilon=0.008 $ \cite{ref11} then $ r=0.128 $ which is less than BICEP2 released data ($ r=0.20_{-0.05}^{+0.07} $) \cite{ref22} and now a question dawns on the mind: “ Is it possible to eliminate this flaw by considering curvature factor?” In order to answer, lets calculate $ r_n $ using equation (\ref{47}) and (\ref{60}) which results in
\begin{multline}\label{61}
r_n=16\left(\mathfrak{RH} \right) ^{-2\left(\epsilon+\delta \right) }\left[\frac{\Gamma\left(\epsilon +\frac{3}{2} \right) }{\Gamma\left( 2\epsilon +\delta +\frac{3}{2}\right) } \right]^2n^{2\left( \epsilon +\delta\right) }\left(2+\frac{2n}{\sqrt{n^2+1}} \right) ^{-\left(\epsilon +\delta \right) }\\
\times\left(1+\frac{1}{n^2} \right) \left(1-\frac{n^2+1}{n^2\left(n^2-4\right) }\epsilon \right) .
\end{multline}
Besides, one can write
\begin{multline}\label{62}
n^{2\left( \epsilon +\delta\right) }=\left( \mathcal{H}^2\vert_{\tau=0}\right) ^{\epsilon +\delta}=\left( \cot^2\Theta\vert_{\tau=0}\right) ^{\epsilon +\delta}=\left( \cos^2\Theta\vert_{\tau=0}\right) ^{\epsilon +\delta}\left( \sin^2\Theta\vert_{\tau=0}\right) ^{-\left( \epsilon +\delta\right) }\\
=\left(1+ \frac{1}{n^2}\right) ^{-\left( \epsilon +\delta\right) }\left(\mathfrak{RH} \right) ^{2\left(\epsilon+\delta \right) }\epsilon .
\end{multline}
On the other hand, it isn’t hard to show that
\begin{equation}\label{63}
\frac{\Gamma\left( 2\epsilon +\delta +\frac{3}{2}\right) } {\Gamma\left(\epsilon +\frac{3}{2} \right)}= 1+\left(2-\gamma-\ln 2 \right) \left(\epsilon +\delta \right) \simeq \exp \left[\left(2-\gamma-\ln 2 \right) \left(\epsilon +\delta \right)  \right] \simeq \left(2.074 \right) ^{\epsilon +\delta} ,
\end{equation}
where $ \gamma\simeq 0.577 $ is the Euler-Mascheroni constant. In order to derive equation (\ref{63}) one can use the following relation\cite{ref36}
\begin{equation}
\frac{\Gamma\left(x+\epsilon +1 \right) }{\Gamma\left(x+1 \right)}=1+\epsilon \left[ -\gamma +\sum _{n=1}^\infty \left(\frac{1}{n}-\frac{1}{x+n} \right) \right] .
\end{equation}
By inserting equations (\ref{62}) and (\ref{63}) in equation (\ref{61}) we can obtain
\begin{multline}\label{64}
r_n=16\epsilon\,e^{\left(-4+2\gamma +2\ln 2 \right) \left(\epsilon +\delta \right)} \left(1+\frac{1}{n^2} \right)^{1-\left( \epsilon +\delta\right) }\left( 2+\frac{2n}{\sqrt{n^2+1}}\right) ^{-\left( \epsilon +\delta\right) }\\
 \times\left(1-\frac{n^2+1}{n^2\left(n^2-4 \right) }\epsilon \right) .
\end{multline}
For $ n\gg 1  $ equation (\ref{64}) reduces to
\begin{equation}\label{65}
r_{n\gg 1} \simeq 16\epsilon\, e^{\left( - 4 + 2\gamma \right)\left(\epsilon  + \delta\right)}\simeq 16\epsilon \left( 0.58\right) ^{\epsilon  + \delta } .
\end{equation}
By considering $ n_\ast=3 $ as the pivot comoving wave number and $ \epsilon=0.008 $ ,one can show $  r_\ast>16\epsilon $ provided that $ \delta\lesssim 0.03 $ (which is laid in the permitted Planck data\footnote{In the Planck collaboration paper, the slow-roll parameters are $ \epsilon _V $ and $ \eta_V $ which in comparison with our definition $ \epsilon_V=\epsilon $ and $ \eta_V=\epsilon-\delta $.}  \cite{ref11}) , so it may reduce discrepancy between BICEP2 results and slow-rolling inflationary theory to some extent, but it isn't statistically significant to eliminate the flaw completely.  
\section{Conclusion and summary}
\label{sec:5}
In this article we investigated an inflationary model with positive curvature index and calculated the scalar and tensor perturbations power spectra associated to it. For the severe super-Hubble scales (i.e. $ n\gg 1 $) it seems both spectra are completely similar to the spatially flat corresponding case. It is shown that this model yields a natural resolving of the running number problem. We also calculated the tensor-scalar ratio and showed it depends on the wave number of the perturbative modes directly. Furthermore, we showed that it doesn’t seems to mitigate the discrepancy between BICEP2 released results and anticipation of the slow-rolling inflationary model by entering the curvature factor. 


\begin{thebibliography}{99}
\bibitem{ref1}
Guth A H,\emph{Inflationary universe: a possible solution to the horizon and flatness problems}, \emph{Phys.Rev. D} {\bf 23} (1981) 347.

\bibitem{ref2}
Linde A D,\emph{A new inflationary universe scenario: A possible solution of the horizon, flatness, homogeneity, isotropy and primordial monopole problems}, \emph{Phys. Lett. B} {\bf 108}, (1982) 389.

\bibitem{ref3}
Albrecht A and Steinhardt P J,\emph{Cosmology for Grand Unified Theories with Radiatively Induced Symmetry Breaking}, \emph{Phys. Rev. Lett.} {\bf 48}, (1982) 1220. 

\bibitem{ref4}
Guth A H and Pi So-Young,\emph{Fluctuations in the new inflationary universe}, \emph{Phys.Rev.Lett.} {\bf 49} (1982) 1110.

\bibitem{ref5}
Bardeen J M, Steinhardt P J and Turner M H,\emph{Spontaneous creation of almost scale-free density perturbations in an inflationary universe }, \emph{Phys.Rev. D} {\bf 28} (1983) 679.

\bibitem{ref6}
Peter P, and Uzan J P,\emph{Primordial cosmology}, Oxford University Press (2009).

\bibitem{ref7}
Lyth D H and Liddle A R,\emph{The primordial density perturbation: cosmology, inflation and the origin of structure}, Cambridge University Press (2009).

\bibitem{ref8}
Harrison E R,\emph{Fluctuations at the Threshold of Classical Cosmology}, \emph{Phys. Rev. D} {\bf 1}, (1970) 2726.

\bibitem{ref9}
Zel'dovich Y B,\emph{A Hypothesis, Unifying the Structure and the Entropy of the Universe}, \emph{ Mon. Not. R. Astron. Soc.} {\bf 160}, (1972) 1P. 

\bibitem{ref10}
Peebles P J E and Yu J T,\emph{ Primeval Adiabatic Perturbation in an Expanding Universe} , \emph{Astrophys. J.} {\bf 162}, (1970) 815.

\bibitem{ref11}
Ade P A R et al.,\emph{Planck 2013 results. XXII. Constraints on inflation}, arXiv:1303.5082.

\bibitem{ref12}
Hinshaw G et al.,\emph{NINE-YEAR WILKINSON MICROWAVE ANISOTROPY PROBE (WMAP) OBSERVATIONS: COSMOLOGICAL PARAMETER RESULTS}, \emph{Astrophys. J. Suppl. Ser.} {\bf 208}, (2013) 19.

\bibitem{ref13}
Weinberg S,\emph{Cosmology},Oxford University Press (2008).

\bibitem{ref14}
Steinhardt P J and Turner M S,\emph{Prescription for successful new inflation}, \emph{Phys. Rev. D} {\bf 29}, (1984) 2162.

\bibitem{ref15}
Salopek D S and Bond J R,\emph{Nonlinear evolution of long-wavelength metric fluctuations in inflationary models}, \emph{Phys. Rev. D} {\bf 42}, (1990) 3936, ; Salopek D S and Bond J R , \emph{Stochastic inflation and nonlinear gravity}, \emph{Phys. Rev. D} {\bf 43}, (1991) 1005.

\bibitem{ref16}
Liddle A R , Parsons P and Barrow J D,\emph{Formalizing the slow-roll approximation in inflation}, \emph{Phys. Rev. D} {\bf 50}, (1994) 7222.

\bibitem{ref17}
Liddle A R and Lyth D H, \emph{COBE, gravitational waves, inflation and extended inflation},\emph{Phys. Lett. B} {\bf 291}, (1992) 391.

\bibitem{ref18}
Olive K A ,\emph{Inflation}, \emph{Phys. Rep.} {\bf 190}, (1990) 307.

\bibitem{ref19}
Stewart E D and Lyth D H,\emph{A more accurate analytic calculation of the spectrum of cosmological perturbations produced during inflation}, \emph{Phys. Lett. B} {\bf 302}, (1993) 171.

\bibitem{ref20}
Lyth D H, \emph{Large-scale energy-density perturbations and inflation},\emph{Phys. Rev. D} {\bf 31}, (1985) 1792.

\bibitem{ref21}
Bardeen J M ,\emph{Gauge-invariant cosmological perturbations}, \emph{Phys. Rev. D} {\bf 22}, (1980) 1882.

\bibitem{ref22}
Ade P A et al.,\emph{BICEP2 I: Detection Of B-mode Polarization at Degree Angular Scales}, arXiv:1403.3985.

\bibitem{ref23}
Tegmark M, de Oliveira-Costa A and Hamilton A J S,\emph{High resolution foreground cleaned CMB map from WMAP}, \emph{Phys. Rev. D} {\bf 68}, (2003) 123523. 

\bibitem{ref24}
Kogut A et al.,\emph{First-Year Wilkinson Microwave Anisotropy Probe (WMAP) Observations: Temperature-Polarization Correlation}, \emph{Astrophys. J. Suppl. Ser.} {\bf 148}, (2003) 161.

\bibitem{ref25}
Uzan J P, Kirchner U and Ellis G F R,\emph{Wilkinson Microwave Anisotropy Probe data and the curvature of space}, \emph{Mon. Not. R. Astron. Soc.} {\bf 344}, (2003) L65.

\bibitem{ref26}
Luminet J P, Weeks J R, Riazuelo A, Lehoucq R and Uzan J P,\emph{Dodecahedral space topology as an explanation for weak wide-angle temperature correlations in the cosmic microwave background}, \emph{Nature} {\bf 425}, (2005) 593.

\bibitem{ref27}
Aurich R, Lustig S and Steiner F,\emph{CMB anisotropy of the Poincaré dodecahedron}, \emph{Class. Quantum Grav.} {\bf 22}, (2005) 2061;  Aurich R, Lustig S and Steiner F,\emph{CMB anisotropy of spherical spaces}, \emph{Class. Quantum Grav.} {\bf 22}, (2005) 3443; Aurich R and Lustig S,\emph{A survey of lens spaces and large-scale cosmic microwave background anisotropy}, \emph{Mon. Not. R. Astron. Soc.} {\bf 424}, (2012) 1556; Aurich R and Lustig S,\emph{Cosmic topology of polyhedral double-action manifolds}, \emph{Class. Quantum Grav.} {\bf 29}, (2012) 235028. 

\bibitem{ref28}
Hawking S W and Ellis G F R, \emph{The Large Scale Structure of Space-Time}, Cambridge University Press (1973).

\bibitem{ref29}
Ade P A R et al.,\emph{Planck 2013 results. XXVI. Background geometry and topology of the Universe}, arXiv:1303.5086.

\bibitem{ref30}
Ellis G F R, Stoeger W, McEwan P and Dunsby P,\emph{Dynamics of Inflationary Universe with Positive Spatial Curvature}, \emph{Gen. Rel. Grav.} {\bf 34}, (2002) 1445.

\bibitem{ref31}
Ellis G F R, McEwan P, Stoeger W and Dunsby P,\emph{Causality in Inflationary Universe with Positive Spatial Curvature}, \emph{Gen. Rel. Grav.} {\bf 34}, (2002) 1461.

\bibitem{ref32}
Vilenkin A,\emph{Birth of inflationary universes}, \emph{Phys. Rev. D} {\bf 27}, (1983) 2848.

\bibitem{ref33}
Vilenkin A,\emph{CREATION OF UNIVERSES FROM NOTHING}, \emph{Phys. Lett. B} {\bf 117}, (1982) 25.

\bibitem{ref34}
Vilenkin A,\emph{Quantum origin of the universe}, \emph{Nucl. Phys. B} {\bf 252}, (1985) 141.

\bibitem{ref35}
Linde A,\emph{Can we have inflation with $ \Omega>1 $ ?}, \emph{J. Cosmol. Astropart. Phys.} {\bf 05} (2003) 002.

\bibitem{ref36}
Ellis G F R and Maartens R,\emph{The emergent universe: inflationary cosmology with no singularity}, \emph{Class. Quantum Grav.} {\bf 21}, (2004) 223.

\bibitem{ref37}
Ellis G F R, Murugan J and Tsagas C G,\emph{The emergent universe: an explicit construction}, \emph{Class. Quantum Grav.} {\bf 21}, (2004) 233.

\bibitem{ref38}
Lasenby A and Doran C, \emph{Closed universes, de Sitter space, and inflation}, \emph{Phys. Rev. D} {\bf 71}, (2005) 063502. 

\bibitem{ref39}
Mukhanov V,\emph{Gravitational instability of the universe filled with a scalar field}, \emph{JETP Lett.} {\bf 41}, (1985) 493. 
 
\bibitem{ref40}
Sasaki M,\emph{Large Scale Quantum Fluctuations in the Inflationary Universe}, \emph{Prog.Theor. Phys.} {\bf 76}, (1986) 1036.
  
\bibitem{ref41}
Asgari A A, Abbassi A H and Khodagholizadeh J,\emph{On the perturbation theory in spatially closed background}, \emph{Eur. Phys. J. C} {\bf 74}, (2014) 2917.

\bibitem{ref42}
Lifshitz E M,\emph{About gravitational stability of expanding worlds}, \emph{J. Phys. (USSR)} {\bf 10}, (1946) 116. 

\bibitem{ref43}
Lifshitz E M and Khalatnikov I M,\emph{Investigations in relativistic cosmology}, \emph{Adv. Phys.} {\bf 12}, (1963) 185.

\bibitem{ref44}
Bunch T S and Davies P C W,\emph{Quantum Field Theory in De Sitter Space: Renormalization by Point-Splitting}, \emph{Proc. R. Soc. A} {\bf 360}, (1978) 117.

\bibitem{ref45}
Lidsey J E et al.,\emph{Reconstructing the inflaton potential—an overview}, \emph{Rev. Mod. Phys.} {\bf 69}, (1997) 373.

\bibitem{ref46}
Abramowitz M and Stegun I A, \emph{Handbook of mathematical functions: with formulas, graphs, and mathematical tables}, Courier Dover Publications (2012).

\bibitem{ref47}
Virchenko N O and Fedotova I, \emph{Generalized associated Legendre functions and their applications}, World Scientific (2001).




\end{thebibliography}
\end{document}